\documentclass[%
 reprint,
 amsmath,amssymb,
 aps,
]{revtex4-2}

\usepackage{graphicx}% Include figure files
\usepackage{dcolumn}% Align table columns on decimal point
\usepackage{bm}% bold math
\begin{document}

\preprint{APS/123-QED}

\title{Isotopic Fingerprints of Proton-mediated Dielectric Relaxation in Solid and Liquid Water}

\author{Alexander Ryzhov}
 \affiliation{Austrian Institute of Technology, 1210 Vienna, Austria}

\author{Pavel Kapralov}
 \affiliation{Israel Institute of Technology, 320000 Haifa, Israel}

 \author{Mikhail Stolov}
 \affiliation{Israel Institute of Technology, 320000 Haifa, Israel}

 \author{Anton Andreev}
 \affiliation{University of Washington, 98195 Seattle, United States}

 \author{Aleksandra Radenovic}
 \affiliation{École Polytechnique Fédérale de Lausanne, 1015 Lausanne, Switzerland}

 \author{Viatcheslav Freger}
 \affiliation{Israel Institute of Technology, 320000 Haifa, Israel}

\author{Vasily Artemov}%
 \email{vasily.artemov@tuhh.de}
\affiliation{École Polytechnique Fédérale de Lausanne, 1015 Lausanne, Switzerland}
\affiliation{Hamburg University of Technology, 21073 Hamburg, Germany}

\date{\today}

\begin{abstract}
We report cross-validated measurements of the isotope effect on dielectric relaxation for four isotopologues of ice and water, including the 1–10$^5$ Hz region, in which only sporadic and inconsistent measurements were previously available. In ice, the relaxation rates exhibit an activated temperature dependence with an isotope-independent activation energy. Across 248–273 K, the H$_2$O/D$_2$O relaxation rate ratio remains constant at 2.0 $\pm$ 0.1. This scaling agrees with Kramers’ theory in the high-friction limit if the moving mass is the proton or deuteron, indicating that dielectric relaxation is governed by a classic proton transfer over an energy barrier rather than molecular reorientation. 
\end{abstract}

\maketitle

%\tableofcontents

%\section{\label{sec:level1}Introduction}

\begin{figure*}
\includegraphics{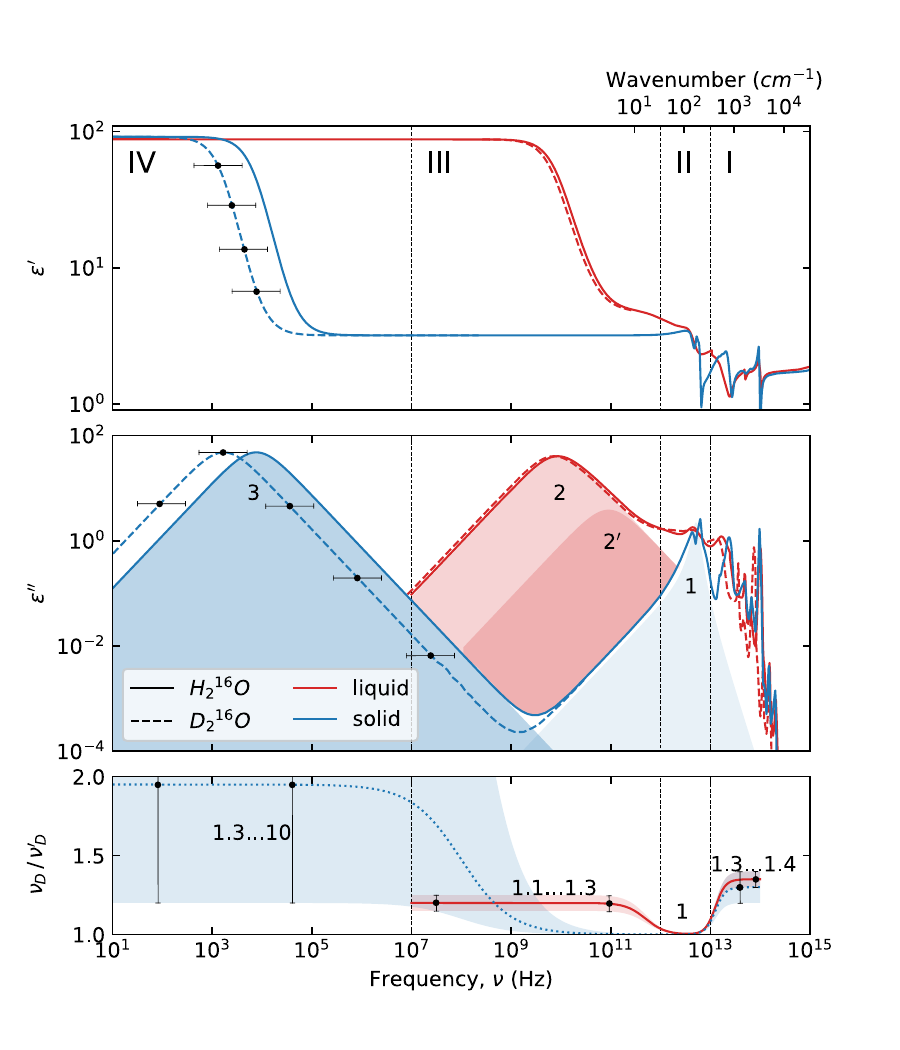}% Here is how to import EPS art
\caption{\label{fig1} \textbf{Generalized dielectric spectra of water (red) and ice (blue)}. Real ($\varepsilon'$) and imaginary ($\varepsilon''$) components of the dielectric function for H$_2$O and D$_2$O, compiled from experimental data \cite{Hor58, Pet99, Art19, Hob74, Hip88, Tak07, War08, Mat96, Lun17, Shi18, Kut21} across four spectral regions: infrared (I), terahertz (II), microwave (III), and radiowave (IV). Key spectral features are: (1) 5-THz vibrational mode; (2, 2$'$) two distinct relaxation modes in water; and (3) a single relaxation mode in ice. The bottom panel shows the isotope effect as a ratio between H$_2$O and D$_2$O relaxation times. Numbers show the isotope effect dispersion in the literature.}
\end{figure*}

Understanding the mechanism of charge dynamics in water and ice is important for atmospheric science, global communications, chemistry, soft matter physics, and biology \cite{Yua25, Kni19, Bal17}. The frequency-dependent dielectric function, $\epsilon^*(\omega) = \epsilon'(\omega) + i\epsilon''(\omega)$, of water and ice that characterizes their electrodynamic properties has been extensively studied over decades \cite{Deb29, Kaa97, Sas16, Lun17, Shi18, Kut21, Ahl22}, and summarized in Fig.~\ref{fig1}. Peaks in the imaginary part, $\epsilon''(\omega)$, identify the characteristic rates of molecular oscillation and relaxation processes, thereby providing insight into the underlying atomic and molecular dynamics. At frequencies above $\sim$1 THz, multiple intramolecular oscillations are observed, which are now well understood. Below 1 THz, the frequency dependence of $\epsilon^*(\omega)$ in ice is well described by just a single Debye relaxation contribution \cite{Deb29}:
\begin{equation}
\varepsilon^*(\omega) = \varepsilon(\infty) + \frac{\Delta \varepsilon_D}{1 + i\omega\tau_D} 
\label{eq:debye}
\end{equation}
where $\varepsilon(\infty)$ is the high-frequency permittivity,  and $\Delta\varepsilon_D$ is the enhancement of permittivity, associated with the relaxation mechanism, and $\tau_D$ is the corresponding relaxation time. The relaxation term shows a maximum in terms of $\epsilon''(\omega)$ at $\omega = 2\pi/\tau_D$. In contrast, liquid water exhibits two distinct relaxation processes: the main and a satellite one. Its dielectric spectrum is therefore well described by a sum of two Debye contributions of the form given by the second term in Eq.~(\ref{eq:debye}). 

%\begin{figure}%[h]
%\centering
%\includegraphics[width=0.5\textwidth]{Fig_1.pdf}

The relaxation peak(s) are associated with the cooperative motion of charges, but the precise physical mechanisms underlying the low-frequency dielectric relaxation and the high polarizability of water and ice are still debated \cite{Hob66, Hip88, Kor03, Art21}. Measuring the isotope shift of these peaks provides a valuable approach for identifying the relaxation mechanism and has been studied extensively \cite{Aut52, Joh76, Kaw79, Oka99}. Both ice and water show a 1.4 D/H ratio for the infrared absorption peaks, indicative of the intramolecular modes. In contrast, the relaxation bands of water display a D/H rate ratio of only about 1.2. Furthermore, in the low-frequency range of 1–10$^5$ Hz of ice, only sporadic and contradictory measurements of the isotope effect are available. This uncertainty, together with the difficulty of modeling this range from first principles, has led to contradictory interpretations ranging from classical molecular reorientation to quantum tunneling \cite{Aut52, Joh81, Sas16, Joh78, Hob74, Bru93, Pet99}. These controversies prevent comparison of the isotope effects of ice and water and, consequently, definitive conclusions about the underlying relaxation mechanism(s).

In this Letter, we report cross-validated measurements of dielectric relaxations in four isotopologues of ice and water across a broad frequency range, including the previously controversial 1–10$^5$ Hz window in which systematic and conclusive data for several ice isotopologues were lacking. Our measurements provide strong evidence for the classical thermally-activated mechanism of dielectric relaxation studied by Kramers, in which the mass of the particles moving over the thermal barrier corresponds to the mass of individual protons/deuterons. This suggests that the low-frequency dielectric relaxation between 248 and 273 K is mediated by the classical motion of protons between different free energy minima, rather than molecular reorientations. 

%\section{Methods}

To probe the dielectric response of liquid and solid water across a broad frequency range, we combined two complementary dielectric spectroscopy techniques \cite{Kre03}, cumulatively spanning a 1–10$^{11}$ Hz range. Low-frequency measurements (1–10$^6$ Hz) were performed with a BioLogic VSP-300 using a custom parallel-plate electrode setup, while high-frequency measurements (10$^5$–10$^{11}$ Hz) were conducted in reflection mode with a Keysight P5008B vector network analyzer and a commercial coaxial probe kit. Ultrapure isotopic water samples (impurities $<$0.1\%) were used as received. Sample handling minimized contact with atmospheric moisture, including airtight sealing of measurement cells and atmosphere-free probe manipulation.

Polycrystalline ice samples were obtained by slow cooling of the liquids within the measurement cells. Temperature was controlled between 248–333 K ($\pm$0.2 K) using a Peltier element integrated into a thermally insulated holder, with heptane coolant circulating around the electrodes. Heating/cooling rates and cell dimensions were optimized to ensure stable and reproducible dielectric responses across the full frequency range. The optimized low-frequency cell had a 7 mm diameter and $\sim$1 mm electrode gap, while the high-frequency cell was sized to suppress boundary reflections and had a diameter of around 2 cm and a depth of around 1 cm.

Raw impedance and power-spectrum data were converted to the dielectric function $\varepsilon^*$ (see Appendix A for details). The real, $\varepsilon'$, and imaginary $\varepsilon''$ components of the latter were fitted using Eq. (1). A cross-validation across the four isotopologues, in both liquid and solid states, resulted in improved accuracy and resolution compared to previous studies.

%\section{Results}

\begin{figure}%[h]
\centering
\includegraphics[width=0.5\textwidth]{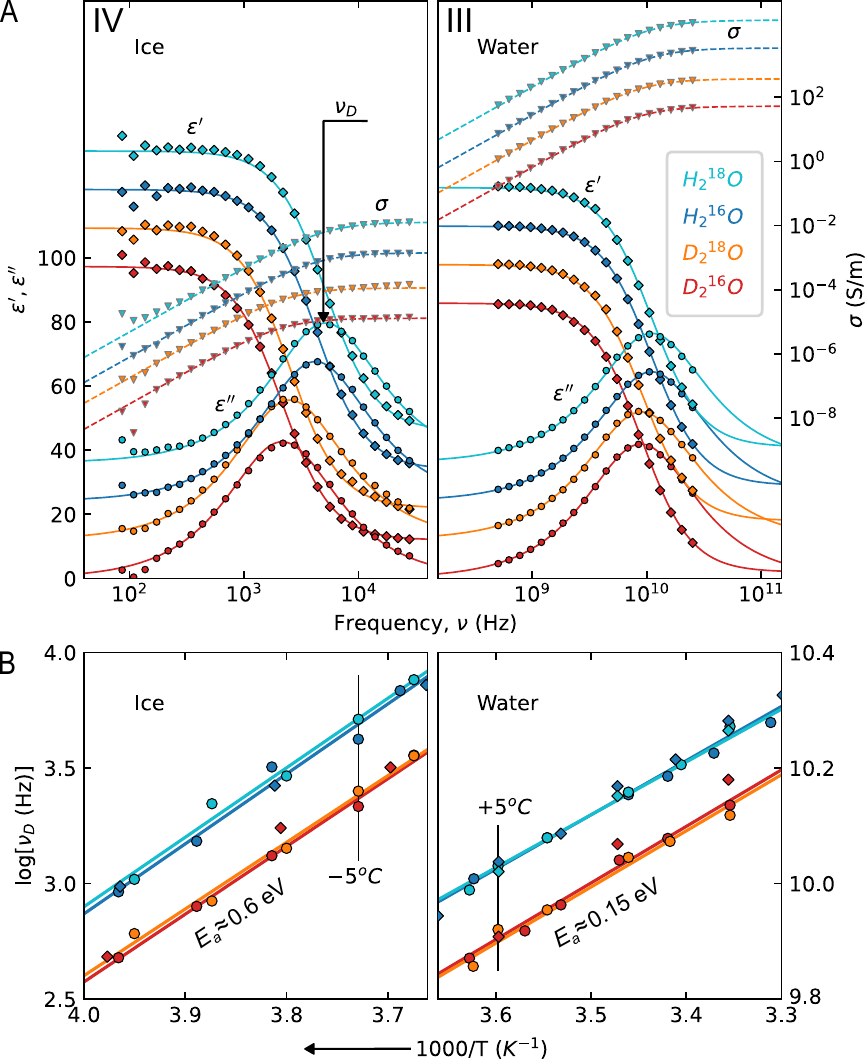}

%\begin{figure*}
%\includegraphics{Fig_2.pdf}
\caption{\textbf{Dielectric function of water and ice isotopologues.} (\textbf{A}) Real and imaginary parts of the dielectric constant for light (H$_2^{16}$O), heavy (D$_2^{16}$O), heavy-oxygen (H$_2^{18}$O), and double-heavy (D$_2^{18}$O) forms of ice (left) and water (right), measured at -5 $^{\circ}$C and +5 $^{\circ}$C, respectively. DC conductivity has been subtracted, and 20 units vertically offset curves for clarity. Solid lines represent best fits to the Debye model. (\textbf{B}) Arrhenius plots showing the temperature dependence of the Debye relaxation frequency $\nu_D$. Vertical lines correspond to the data points in panel A. Circles denote measurements from this work; diamonds indicate previously reported, but incomplete, literature data.}\label{fig2}
\end{figure}

Figure $\ref{fig2}$A shows the dielectric relaxation band of four water isotopologues (see legend) in solid and liquid phases in terms of the real, $\varepsilon'$, and imaginary, $\varepsilon''$, parts of the dielectric permittivity, and the dynamic conductivity, $\sigma = \varepsilon''\varepsilon_02\pi\nu$. The dots are experimental data, and the lines represent a model derived from Eq. (\ref{eq:debye}). Table~\ref{tab:table1} shows the parameters of the main dielectric relaxation bands of ice (superscript $s$) and water (superscript $l$) around the melting point. 

\begin{table}[b]
    \caption{\label{tab:table1}Dielectric parameters of water and ice isotopologues: the permittivity enhancement ($\Delta\varepsilon_D$), the Debye relaxation frequency ($\nu_D = \tau_D^{-1}$) in Hz, and high-frequency (ac) conductivity plateau ($\sigma_{ac}$) in S/m. All parameters are given at 273 $\pm$ 5 K\footnote{Superscripts $l$ and $s$ are for liquid and solid, respectively}.}
    \begin{ruledtabular}
       \begin{tabular} {ccccccc}
         & $\Delta\varepsilon^l_D$ & $\Delta\varepsilon^s_D$ & $\nu_D^l$ & $\nu_D^s$ & $\sigma_{ac}^l$ & $\sigma_{ac}^s$ \\
         \hline
         H$_2$O$^{16}$ & 85 & 98 & 10.7$\cdot 10^{9}$ & 4.9$\cdot 10^{3}$ & 62.5 & 2.6$\cdot 10^{-5}$ \\
         D$_2$O$^{16}$ & 86 & 96 & 8.1$\cdot 10^{9}$ & 2.3$\cdot 10^{3}$ & 52.5 & 1.3$\cdot 10^{-5}$ \\
         H$_2$O$^{18}$ & 83 & 98 & 10.8$\cdot 10^{9}$ & 5.2$\cdot 10^{3}$ & 62.6 & 3.2$\cdot 10^{-5}$ \\
         D$_2$O$^{18}$ & 84 & 99 & 7.9$\cdot 10^{9}$ & 2.4$\cdot 10^{3}$ & 50.6 & 1.5$\cdot 10^{-5}$ \\
       \end{tabular}
\end{ruledtabular}
\end{table}

Figure $\ref{fig2}$B shows the temperature dependence of the relaxation frequency, $\nu_D=\tau_I^{-1}$. Despite the six orders of magnitude discontinuity at the melting point, both ice and water follow perfect Arrhenius behavior, $\nu_D(T) = A \exp(-E_a / k_B T)$, between 248 and 333 K. The activation energies are 0.57 $\pm$ 0.01 eV for ice and 0.15 $\pm$ 0.01 eV for water, with a ratio of about 3.8. Both values are an order of magnitude larger than the thermal energy $k_BT \approx 0.02$ eV, reflecting a substantial barrier to relaxation. Notably, $E_a$ is invariant under isotope substitution, which alters only the prefactor $A$, making the isotope effect simply the ratio $A/A'$ of prefactors for two isotopologues at the same temperature.

\begin{figure}%[h]
\centering
\includegraphics[width=0.48\textwidth]{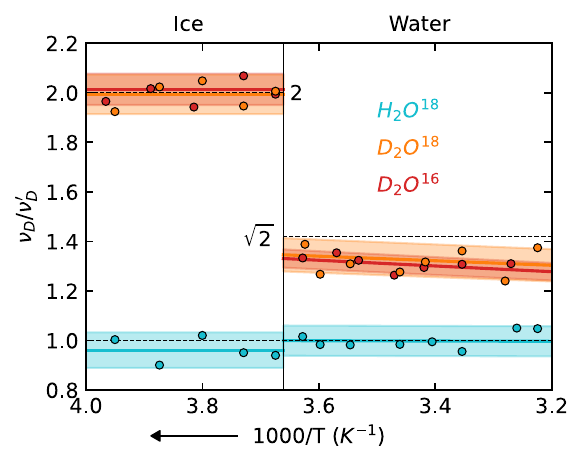}
\caption{\textbf{Isotope effect on the dielectric relaxation time in ice and water.} The data points show the ratio of the relaxation time of three isotopologues against the H$_2$O$^{16}$ (normal water). Dashed lines are guides for the eyes and correspond to the isotopic ratio of 1 (no isotope effect), $\sqrt{2}$ (rotational diffusion), and 2 (proton hopping).}\label{fig3}
\end{figure}

Our data reveal no detectable isotope effect in either ice or water upon $^{16}$O$\,\to\,^{18}$O substitution, whereas replacing $^1$H with $^2$H significantly alters $\nu_D$ in both light- and heavy-oxygen variants (Fig.~\ref{fig3}). Note that isotope H$_2^{16}$O has served as the reference. The relaxation-time ratio between light and heavy water is non-monotonic at the phase transition. In liquid water, the isotope effect of $\approx$1.2 is slightly temperature dependent, that consistent with previous reports \cite{Lun17, Shi18, Kut21}, while in ice, the proton/deuteron substitution yields a constant ratio of 2.0 $\pm$ 0.1 between 248 and 273 K, and has not been explicitly reported before.

%\section*{Discussion}

Although the temperature dependence of the dielectric constants of ice and water reported elsewhere \cite{Art14, Art19} is nearly smooth across the transition between the two phases, the isotope effect exhibits a discontinuity, pointing to a different relaxation mechanism. Neither of them can be explained within the simplified picture of rotational dynamics of water molecules because the found isotope ratio in water is smaller than the $\sqrt{2} \approx 1.4$ expected from dipole reorientation, whereas in ice it is notably larger.

\begin{figure}%[h]
\centering
\includegraphics[width=0.30\textwidth]{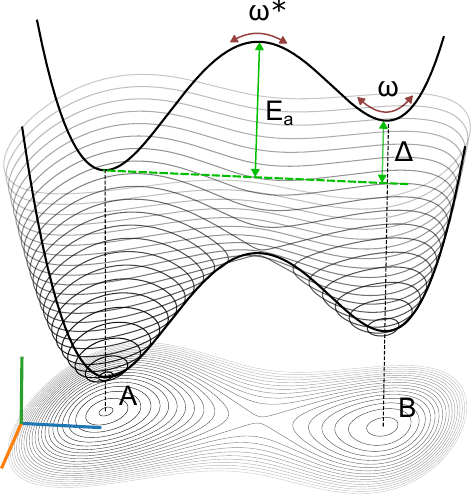}
\caption{\textbf{Energy landscape of the dielectric relaxation}. States $A$ and $B$ correspond to the initial and final states, respectively, separated by an activation barrier $E_a$ and determining the free energy difference $\Delta$. Frequencies $\omega^*$ and $\omega$ are the transition frequency and the attempt frequency, respectively.}
\label{fig4}
\end{figure}

The Arrhenius behavior of the relaxation time indicates a \textit{classical} barrier-crossing mechanism between two minima (Fig.~\ref{fig4}), consistent with Kramers' theory \cite{Kra40}. In the high-friction regime ($\eta \gg \omega^{*}$), this transition rate theory gives a prefactor:
\begin{equation}
  A = \frac{2\pi\, \omega\, \omega^{*}}{\eta},
  \label{eq:kramers_prefactor}
\end{equation}
where $\omega=\sqrt{\kappa/m}$ and $\omega^{*}=\sqrt{\kappa^{*}/m^{*}}$ are the characteristic frequencies near the potential minimum and barrier top, respectively, and $\eta$ is the (dynamic) friction coefficient. Isotope effects on $\eta$, which differ from the static macroscopic viscosity, are negligible to first order \cite{Han90}, and the isotope-independence of the activation energy $E_a$ implies that $\kappa$ and $\kappa^{*}$ are also isotope independent.

The isotope effect in Kramers' theory depends on the mass involved in the motion at the potential minimum, $m$, and the top of the barrier, $m^*$, and for water molecules takes the following three values depending on the choice:

\begin{widetext}
\begin{equation}
\frac{\nu_D}{\nu'_D} = \sqrt{\frac{m' m'^*}{m m^*}} \cdot \frac{\eta'}{\eta} \approx 
\begin{cases}
2.0, & \text{both masses are hydrogens}, \\
1.4, & \text{one mass is hydrogen, another includes oxygen}, \\
1.1, & \text{both masses include oxygen}.
\end{cases}
\label{eq:high-friction}
\end{equation}
\end{widetext}
Comparison of the coefficients in Eq. (\ref{eq:high-friction}) with experimental data supports a model in which proton hopping (case 1), rather than molecular reorientations (case 2) or collective molecular rearrangements (case 3), governs dielectric relaxation in ice. However, the nature of the states $A$ and $B$, connected by a single proton transfer, cannot be determined unambiguously from experiment.

Within the standard picture of ice as an oxygen lattice with mobile protons obeying the ice rules \footnote{Each O–O bond hosts exactly one proton, displaced from the midpoint so that each oxygen atom is coordinated by two near and two distant protons, giving rise to residual entropy.}, two main candidates that violate these rules emerge. The first are Bjerrum defects \footnote{Bjerrum defects are a bond without a proton (L-defect) or with two protons (D-defect).}, which form through molecular rotations and involve intramolecular proton transfer. The second are Bjerrum pairs, arising from proton transfer between neighboring molecules to produce closely spaced ions \footnote{Bjerrum argued [Bjerrum, N. Untersuchungen über Ionenassoziation I. K. Dan. Vidensk. Selsk. 1926, 7, 1–48] that under normal conditions thermal energy is insufficient to overcome Coulomb attraction, so ions should form pairs rather than remain free.}. These two processes differ in their isotope signatures. Formation of orientational defects requires an intramolecular proton transfer around the oxygen nucleus, which is likely accompanied by coordinated motion of the oxygen atom. This reduces the potential barrier and should yield an isotope shift smaller than 2. In contrast, intermolecular proton transfer leading to Bjerrum ion pairs does not necessarily involve oxygen motion. Based on this, we attribute the low-frequency relaxation peak in ice to the dissociation and recombination of closely spaced Bjerrum pairs. These pairs have long been discussed in liquid water \cite{Gei01, Cer13, Art19}, and given the small temperature difference between ice and water, and the shared mechanism of molecular dissociation via intermolecular proton transfer, we conjecture that Bjerrum pairs should also be abundant in ice.

In this picture, minimum $A$ in Fig.~\ref{fig4} corresponds to a proton localized within a neutral molecule, while minimum $B$ represents a transient dissociation state. The relaxation time is then linked to the recombination dynamics of short-lived H$_3$O$^+$ and OH$^-$ ions, mediated by proton hopping across the activation barrier. It is easy to show that within this phenomenological model, the dielectric response follows Eq. (\ref{eq:debye}) with a single relaxation time $\tau$ (see Appendix B for details). Importantly, however, the relaxation rate $\tau^{-1}$ reflects the proton-escape dynamics analyzed by Kramers \cite{Kra40}, rather than Debye’s model of orientation of non-interacting dipoles \cite{Deb29} or Onsager’s description of collective rearrangements of correlated dipolar species \cite{Ons36}.

%\section{Conclusions}

In conclusion, we found that the isotope effect on the dielectric relaxation of ice has a factor of 2.0 $\pm$ 0.1, a value not explicitly reported earlier. This finding points to a proton-hopping relaxation mechanism distinct from molecular reorientation. This assumes a self-consistent phenomenological model of ice disorder, pointing to short-lived ionic pairs, analogous to those proposed in aqueous electrolytes (Bjerrum pairs). The transient Bjerrum pairs may contribute to long-range correlations in ice, and presumably in water. Although the factor 2 isotope shift of the low-frequency peak in ice was not previously reported in the literature, a close inspection of the data \cite{Gei01, Cer13} points to numerous short-lived hydronium and hydroxide ions and confirms our conclusions. We anticipate that our results will stimulate further experimental and theoretical efforts to resolve the Bjerrum pairs in solid and liquid water.

\begin{acknowledgments}
We thank Boris Spivak, Alexander Schlaich, Hugo Dil, and Patrick Huber for fruitful discussions. We acknowledge the Centre for Molecular Water Science and the research cluster BlueMat for close collaboration on related topics.

%\section*{Funding}

M.S. and V.F.  acknowledge the financial support from the Israel Science Foundation Grant No. 486/22. V.A. acknowledges the EPFL vice presidency for innovation for the Ignition grant. The work of A.A. was supported by the NSF grant DMR-2424364, the Thouless Institute for Quantum Matter (TIQM), and the
College of Arts and Sciences at the University of Washington.
\end{acknowledgments}

\appendix
\section{Data processing protocol}

\begin{figure}
\centering
\includegraphics[width=1.0\linewidth]{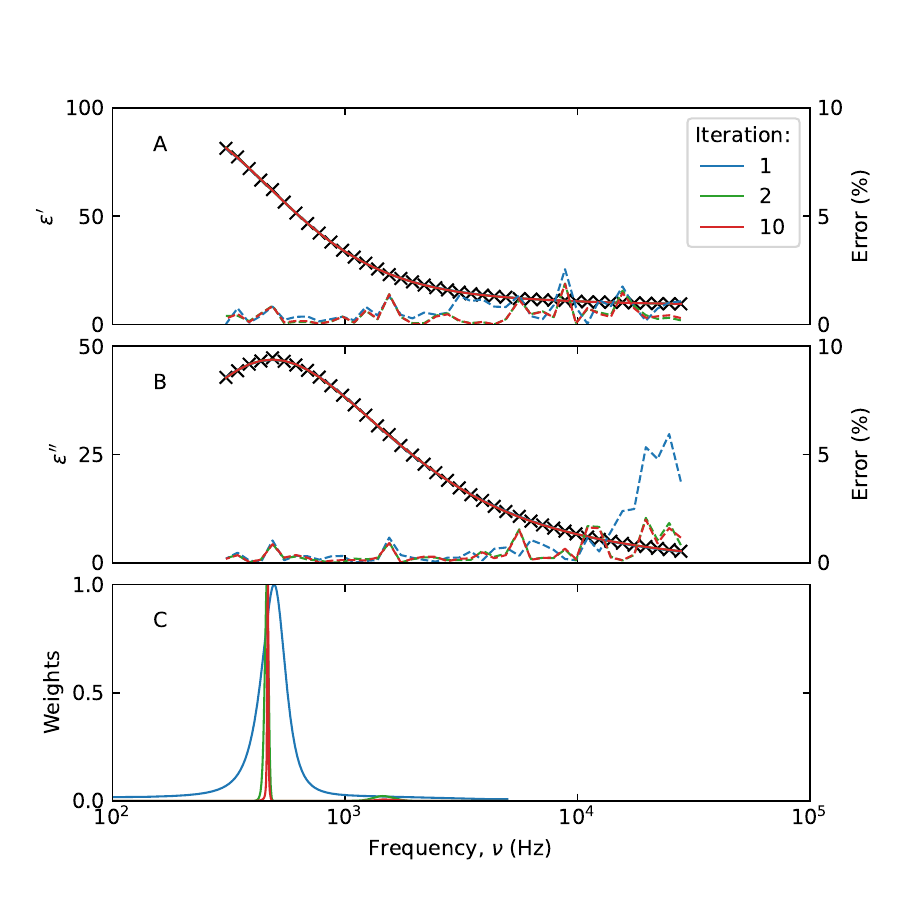}
\caption{\textbf{Dielectric data processing.} (A) Real and (B) imaginary parts of the dielectric permittivity. Symbols ($\times$) represent experimental data, solid lines are fits using Eq.~(S1), and dashed lines show relative fitting error. (C) Evolution of Debye mode weights after the 1st, 2nd, and 10th iterations, demonstrating convergence of the fitting procedure.}
\label{fig:data}
\end{figure}

The frequency dependence of the real ($Z'$) and imaginary ($Z''$) parts of the measured impedance was converted to the dimensionless real ($\epsilon'$) and imaginary ($\epsilon''$) parts of the dielectric permittivity (Fig.~\ref{fig:data}A,B) using standard formalism \cite{Kre03} and the geometry of the measuring cell. The resulting spectra for all isotopologues, in both liquid and solid states and at different temperatures, were fitted with the following model function:

\begin{eqnarray}
\epsilon(\omega) = \epsilon'(\omega) + i\epsilon''(\omega)
= \epsilon_{\infty} + \frac{A_{\mathrm{W}}}{(1+i)\omega^{0.5}}\nonumber\\ + i\frac{\sigma_{0}}{\epsilon_0\omega} 
+ \Delta\epsilon \sum_i \frac{w_i \cdot \omega_{D,i}}{\omega_{D,i} + i\omega}. 
\end{eqnarray}
Here, $\epsilon_\infty$ is the dielectric constant, incorporating dielectric contributions at frequencies higher than fitted, $A_{\mathrm{W}}$ is the Warburg coefficient describing electrode polarization, and the third term accounts for the dc conductivity $\sigma_{0}$ at $\omega \rightarrow 0$. The final term represents a set of Debye relaxations with characteristic frequencies $\omega_{D,i}$ and relative weights $w_i$, where $\Delta\epsilon = \epsilon(0) - \epsilon(\infty)$.

To fit the model to the experimental data, we used the quasi-Newton Broyden–Fletcher–Goldfarb–Shanno (BFGS) algorithm \cite{Fle87}. Free parameters were $\epsilon_\infty$, $A_{\mathrm{W}}$, $\sigma_{0}$, $\Delta\epsilon$, and the weights $w_i$. The Debye frequencies $\omega_{D,i} = 2\pi\nu_{D,i}$ were initialized from the maxima of the relaxation peaks in $\epsilon''(\omega)$ and kept fixed during fitting. To impose the normalization constraint $\sum_i w_i = 1$, the weights were expressed in softmax form:

\begin{equation}
    w_i = \frac{\exp(\tilde{w}_i)}{\sum_j \exp(\tilde{w}_j)} .
\end{equation}

The fitting procedure was performed iteratively. After each iteration, modes with $w_i$ below a threshold were discarded, and the remaining weights were renormalized. This procedure sharpened the dominant relaxation peaks and improved stability of the decomposition (Fig.~\ref{fig:data}C). Convergence was typically achieved within 5–10 iterations. For presentation of the final spectra (main text Fig.~\ref{fig2}), the dc conductivity contribution and electrode polarization ($A_{\mathrm{W}}$) were subtracted.

This protocol provides reliable dielectric spectra without the need for an external reference. Cross-validation among isotopologues enhances accuracy and enables the determination of isotope shifts with higher precision than would be possible from independent measurements alone.

\section{Debye relaxation from generation-recombination of ionic pairs}

We show that the dielectric relaxation arising from the generation and recombination of short-lived intrinsic ice's ionic defects (Bjerrum pairs, BPs) is equivalent to the Debye formula [main text, Eq.~(1)], with the relaxation time $\tau$ corresponding to the characteristic timescale of the ionization equilibrium of Bjerrum ionic pairs.  

Because of their relatively low concentration, BPs in ice can be treated as a dilute solid solution in the crystalline host. In the lattice, BPs with different dipole moment orientations occupy distinct potential minima separated by large barriers. Thus, BPs with different orientations can be regarded as distinct solute species. 

We assume an external electric field $E$ applied along the principal axis of the ice crystal, and adopt a uniaxial (Ising-type) model with two types of BPs: those oriented parallel ($+$) and antiparallel ($-$) to $E$. Following the standard treatment of dilute solutions (see, e.g., \S87 and Eq.~(87.5) in Ref.~\cite{Lan80}), the chemical potential of a BP in an external field $E$ is  
\begin{equation}
\label{eq:mu_BP}
\mu'_{\pm} = kT \ln c_\pm + \psi \mp Ed,
\end{equation}
where $c_\pm \ll 1$ are the concentrations of $+$ and $-$ BPs, $d$ is the dipole increment per BP, and $\psi$ is a thermodynamic term depending on temperature and pressure (or more generally, anisotropic stress in ice), but independent of concentration.  

Since we measure the chemical potential of BPs relative to that of water molecules, the condition for equilibrium generation/recombination of BPs is $\mu_\pm = 0$. This gives the equilibrium BP concentrations for a time-dependent field $E(t)$:  
\begin{equation}
\label{eq:c_pm_equilibrium}
c^{(0)}_\pm (E(t)) = e^{-\psi / kT} \, e^{\pm Ed/kT}
\approx c^{(0)} \left(1 \pm \frac{E(t)d}{kT}\right),
\end{equation}
where $c^{(0)} = e^{-\psi / kT}$ is the equilibrium concentration of each BP species at $E=0$.  

%In the presence of a time-dependent field $E(t)$, the actual BP concentrations deviate from instantaneous equilibrium. In linear response, these deviations are small and obey the relaxation equation:  
%\begin{equation}
%\label{eq:relaxation_eq}
%\dot c_\pm (t) = - \frac{c_\pm(t) - c^{(0)}_\pm (E(t))}{\tau}.
%\end{equation}

Using the Fourier convention $E(t)=Re[E_\omega e^{+i\omega t}]$ and the following sequence of substitutions:  

\begin{align}
& E(t) = Re\big[ E_\omega e^{+i\omega t} \big], 
\qquad
c^{(0)}_\pm(E) \approx c^{(0)}\Big(1 \pm \frac{E d}{kT}\Big), \\[6pt]
& \dot c_\pm(t) = -\frac{c_\pm(t)-c^{(0)}_\pm(E(t))}{\tau}
\quad\Rightarrow\quad \nonumber\\
&c_\pm(t) = c^{(0)} \Big(1 \pm Re\big[ \tfrac{E_\omega d}{kT(1 + i\omega\tau)} e^{+i\omega t} \big]\Big), \\[6pt]
& c_+(t)-c_-(t) = 2 c^{(0)} Re\big[ \tfrac{E_\omega d}{kT(1 + i\omega\tau)} e^{+i\omega t} \big], \\[6pt]
& P(t) = n_w d\big(c_+(t)-c_-(t)\big)= \nonumber\\
&      = 2 c^{(0)} n_w \, Re\big[ \tfrac{E_\omega d^2}{kT(1 + i\omega\tau)} e^{+i\omega t} \big],
\end{align}
where $n_w$ is the number density of water molecules, the complex amplitude of the polarization is
\[
P_\omega = \frac{2 c^{(0)} n_w d^2}{kT(1 + i\omega\tau)}\,E_\omega .
\]

From ${\bf D} = \epsilon_0 {\bf E} + {\bf P}$ and $D_\omega=\epsilon_0\epsilon(\omega)E_\omega$, we obtain
\begin{equation}
\label{eq:dielectric_function}
\epsilon(\omega) = \epsilon (\infty) + \frac{2 c^{(0)} n_w d^2}{\epsilon_0 kT \,(1 + i\omega\tau)}.
\end{equation}
Here $\epsilon (\infty) \approx 3$ is the high-frequency dielectric constant (plateau above $\sim$5 kHz). The factor of 2 arises from the two BP species ($+$ and $-$).

Equation (B7) corresponds to the Debye relaxation formula given by Eq.(1) of the main text.

\bibliography{apssamp}% Produces the bibliography via BibTeX.

\begin{thebibliography}{40}%
\makeatletter
\providecommand \@ifxundefined [1]{%
 \@ifx{#1\undefined}
}%
\providecommand \@ifnum [1]{%
 \ifnum #1\expandafter \@firstoftwo
 \else \expandafter \@secondoftwo
 \fi
}%
\providecommand \@ifx [1]{%
 \ifx #1\expandafter \@firstoftwo
 \else \expandafter \@secondoftwo
 \fi
}%
\providecommand \natexlab [1]{#1}%
\providecommand \enquote  [1]{``#1''}%
\providecommand \bibnamefont  [1]{#1}%
\providecommand \bibfnamefont [1]{#1}%
\providecommand \citenamefont [1]{#1}%
\providecommand \href@noop [0]{\@secondoftwo}%
\providecommand \href [0]{\begingroup \@sanitize@url \@href}%
\providecommand \@href[1]{\@@startlink{#1}\@@href}%
\providecommand \@@href[1]{\endgroup#1\@@endlink}%
\providecommand \@sanitize@url [0]{\catcode `\\12\catcode `\$12\catcode `\&12\catcode `\#12\catcode `\^12\catcode `\_12\catcode `\%12\relax}%
\providecommand \@@startlink[1]{}%
\providecommand \@@endlink[0]{}%
\providecommand \url  [0]{\begingroup\@sanitize@url \@url }%
\providecommand \@url [1]{\endgroup\@href {#1}{\urlprefix }}%
\providecommand \urlprefix  [0]{URL }%
\providecommand \Eprint [0]{\href }%
\providecommand \doibase [0]{https://doi.org/}%
\providecommand \selectlanguage [0]{\@gobble}%
\providecommand \bibinfo  [0]{\@secondoftwo}%
\providecommand \bibfield  [0]{\@secondoftwo}%
\providecommand \translation [1]{[#1]}%
\providecommand \BibitemOpen [0]{}%
\providecommand \bibitemStop [0]{}%
\providecommand \bibitemNoStop [0]{.\EOS\space}%
\providecommand \EOS [0]{\spacefactor3000\relax}%
\providecommand \BibitemShut  [1]{\csname bibitem#1\endcsname}%
\let\auto@bib@innerbib\@empty
%</preamble>
\bibitem [{\citenamefont {Yuan}\ \emph {et~al.}(2025)\citenamefont {Yuan}, \citenamefont {Li}, \citenamefont {Zhu}, \citenamefont {Qiao}, \citenamefont {Kang}, \citenamefont {Wang}, \citenamefont {Tian}, \citenamefont {Huang},\ and\ \citenamefont {Lai}}]{Yua25}%
  \BibitemOpen
  \bibfield  {author} {\bibinfo {author} {\bibfnamefont {Y.}~\bibnamefont {Yuan}}, \bibinfo {author} {\bibfnamefont {J.}~\bibnamefont {Li}}, \bibinfo {author} {\bibfnamefont {Y.}~\bibnamefont {Zhu}}, \bibinfo {author} {\bibfnamefont {Y.}~\bibnamefont {Qiao}}, \bibinfo {author} {\bibfnamefont {Z.}~\bibnamefont {Kang}}, \bibinfo {author} {\bibfnamefont {Z.}~\bibnamefont {Wang}}, \bibinfo {author} {\bibfnamefont {X.}~\bibnamefont {Tian}}, \bibinfo {author} {\bibfnamefont {H.}~\bibnamefont {Huang}},\ and\ \bibinfo {author} {\bibfnamefont {W.}~\bibnamefont {Lai}},\ }\bibfield  {title} {\bibinfo {title} {Water in electrocatalysis},\ }\href {https://doi.org/10.1002/anie.202425590} {\bibfield  {journal} {\bibinfo  {journal} {Angewandte Chemie International Edition}\ }\textbf {\bibinfo {volume} {64}},\ \bibinfo {pages} {e202425590} (\bibinfo {year} {2025})}\BibitemShut {NoStop}%
\bibitem [{\citenamefont {Knight}\ \emph {et~al.}(2019)\citenamefont {Knight}, \citenamefont {Kalugin}, \citenamefont {Coker},\ and\ \citenamefont {Ilgen}}]{Kni19}%
  \BibitemOpen
  \bibfield  {author} {\bibinfo {author} {\bibfnamefont {A.~W.}\ \bibnamefont {Knight}}, \bibinfo {author} {\bibfnamefont {N.~G.}\ \bibnamefont {Kalugin}}, \bibinfo {author} {\bibfnamefont {E.}~\bibnamefont {Coker}},\ and\ \bibinfo {author} {\bibfnamefont {A.~G.}\ \bibnamefont {Ilgen}},\ }\bibfield  {title} {\bibinfo {title} {Water properties under nano-scale confinement},\ }\href {https://doi.org/10.1038/s41598-019-44651-z} {\bibfield  {journal} {\bibinfo  {journal} {Scientific Reports}\ }\textbf {\bibinfo {volume} {9}},\ \bibinfo {pages} {8246} (\bibinfo {year} {2019})}\BibitemShut {NoStop}%
\bibitem [{\citenamefont {Ball}(2017)}]{Bal17}%
  \BibitemOpen
  \bibfield  {author} {\bibinfo {author} {\bibfnamefont {P.}~\bibnamefont {Ball}},\ }\bibfield  {title} {\bibinfo {title} {Water is an active matrix of life for cell and molecular biology},\ }\href {https://doi.org/10.1073/pnas.1703781114} {\bibfield  {journal} {\bibinfo  {journal} {Proceedings of the National Academy of Sciences}\ }\textbf {\bibinfo {volume} {114}},\ \bibinfo {pages} {13327} (\bibinfo {year} {2017})}\BibitemShut {NoStop}%
\bibitem [{\citenamefont {Debye}(1929)}]{Deb29}%
  \BibitemOpen
  \bibfield  {author} {\bibinfo {author} {\bibfnamefont {P.}~\bibnamefont {Debye}},\ }\href@noop {} {\emph {\bibinfo {title} {Polar Molecules}}}\ (\bibinfo  {publisher} {The Chemical Catalogue Company},\ \bibinfo {address} {New York},\ \bibinfo {year} {1929})\BibitemShut {NoStop}%
\bibitem [{\citenamefont {Kaatze}(1997)}]{Kaa97}%
  \BibitemOpen
  \bibfield  {author} {\bibinfo {author} {\bibfnamefont {U.}~\bibnamefont {Kaatze}},\ }\bibfield  {title} {\bibinfo {title} {The dielectric properties of water in its different states of interaction},\ }\href {https://doi.org/10.1007/BF02768829} {\bibfield  {journal} {\bibinfo  {journal} {Journal of Solution Chemistry}\ }\textbf {\bibinfo {volume} {26}},\ \bibinfo {pages} {1049} (\bibinfo {year} {1997})}\BibitemShut {NoStop}%
\bibitem [{\citenamefont {Sasaki}\ \emph {et~al.}(2016)\citenamefont {Sasaki}, \citenamefont {Kita}, \citenamefont {Shinyashiki},\ and\ \citenamefont {Yagihara}}]{Sas16}%
  \BibitemOpen
  \bibfield  {author} {\bibinfo {author} {\bibfnamefont {K.}~\bibnamefont {Sasaki}}, \bibinfo {author} {\bibfnamefont {R.}~\bibnamefont {Kita}}, \bibinfo {author} {\bibfnamefont {N.}~\bibnamefont {Shinyashiki}},\ and\ \bibinfo {author} {\bibfnamefont {S.}~\bibnamefont {Yagihara}},\ }\bibfield  {title} {\bibinfo {title} {Dielectric relaxation time of ice-{I}h with different preparation},\ }\href {https://doi.org/10.1021/acs.jpcb.6b01218} {\bibfield  {journal} {\bibinfo  {journal} {J. Phys. Chem. B}\ }\textbf {\bibinfo {volume} {120}},\ \bibinfo {pages} {3950} (\bibinfo {year} {2016})}\BibitemShut {NoStop}%
\bibitem [{\citenamefont {Lunkenheimer}\ \emph {et~al.}(2017)\citenamefont {Lunkenheimer}, \citenamefont {Emmert}, \citenamefont {Gulich}, \citenamefont {K{\"o}hler}, \citenamefont {Wolf}, \citenamefont {Schwab},\ and\ \citenamefont {Loidl}}]{Lun17}%
  \BibitemOpen
  \bibfield  {author} {\bibinfo {author} {\bibfnamefont {P.}~\bibnamefont {Lunkenheimer}}, \bibinfo {author} {\bibfnamefont {S.}~\bibnamefont {Emmert}}, \bibinfo {author} {\bibfnamefont {R.}~\bibnamefont {Gulich}}, \bibinfo {author} {\bibfnamefont {M.}~\bibnamefont {K{\"o}hler}}, \bibinfo {author} {\bibfnamefont {M.}~\bibnamefont {Wolf}}, \bibinfo {author} {\bibfnamefont {M.}~\bibnamefont {Schwab}},\ and\ \bibinfo {author} {\bibfnamefont {A.}~\bibnamefont {Loidl}},\ }\bibfield  {title} {\bibinfo {title} {Electromagnetic-radiation absorption by water},\ }\href {https://doi.org/10.1103/PhysRevE.96.062607} {\bibfield  {journal} {\bibinfo  {journal} {Physical Review E}\ }\textbf {\bibinfo {volume} {96}},\ \bibinfo {pages} {062607} (\bibinfo {year} {2017})}\BibitemShut {NoStop}%
\bibitem [{\citenamefont {Shiraga}\ \emph {et~al.}(2018)\citenamefont {Shiraga}, \citenamefont {Tanaka}, \citenamefont {Arikawa}, \citenamefont {Saito},\ and\ \citenamefont {Ogawa}}]{Shi18}%
  \BibitemOpen
  \bibfield  {author} {\bibinfo {author} {\bibfnamefont {K.}~\bibnamefont {Shiraga}}, \bibinfo {author} {\bibfnamefont {K.}~\bibnamefont {Tanaka}}, \bibinfo {author} {\bibfnamefont {T.}~\bibnamefont {Arikawa}}, \bibinfo {author} {\bibfnamefont {S.}~\bibnamefont {Saito}},\ and\ \bibinfo {author} {\bibfnamefont {Y.}~\bibnamefont {Ogawa}},\ }\bibfield  {title} {\bibinfo {title} {Reconsideration of the relaxational and vibrational line shapes of liquid water based on ultrabroadband dielectric spectroscopy},\ }\href {https://doi.org/10.1039/c8cp04778b} {\bibfield  {journal} {\bibinfo  {journal} {Physical Chemistry Chemical Physics}\ }\textbf {\bibinfo {volume} {20}},\ \bibinfo {pages} {26200} (\bibinfo {year} {2018})}\BibitemShut {NoStop}%
\bibitem [{\citenamefont {Kutus}\ \emph {et~al.}(2021)\citenamefont {Kutus}, \citenamefont {Shalit}, \citenamefont {Hamm},\ and\ \citenamefont {Hunger}}]{Kut21}%
  \BibitemOpen
  \bibfield  {author} {\bibinfo {author} {\bibfnamefont {B.}~\bibnamefont {Kutus}}, \bibinfo {author} {\bibfnamefont {A.}~\bibnamefont {Shalit}}, \bibinfo {author} {\bibfnamefont {P.}~\bibnamefont {Hamm}},\ and\ \bibinfo {author} {\bibfnamefont {J.}~\bibnamefont {Hunger}},\ }\bibfield  {title} {\bibinfo {title} {Dielectric response of light, heavy and heavy oxygen water: isotope effects on the hydrogen bonding network’s collective relaxation dynamics},\ }\href {https://doi.org/10.1039/d0cp06460b} {\bibfield  {journal} {\bibinfo  {journal} {Phys. Chem. Chem. Phys.}\ }\textbf {\bibinfo {volume} {23}},\ \bibinfo {pages} {5467} (\bibinfo {year} {2021})}\BibitemShut {NoStop}%
\bibitem [{\citenamefont {Ahlmann}\ \emph {et~al.}(2022)\citenamefont {Ahlmann}, \citenamefont {Hoffmann}, \citenamefont {Keppler}, \citenamefont {Münzner}, \citenamefont {Tonauer}, \citenamefont {Loerting}, \citenamefont {Gainaru},\ and\ \citenamefont {Böhmer}}]{Ahl22}%
  \BibitemOpen
  \bibfield  {author} {\bibinfo {author} {\bibfnamefont {S.}~\bibnamefont {Ahlmann}}, \bibinfo {author} {\bibfnamefont {L.}~\bibnamefont {Hoffmann}}, \bibinfo {author} {\bibfnamefont {M.}~\bibnamefont {Keppler}}, \bibinfo {author} {\bibfnamefont {P.}~\bibnamefont {Münzner}}, \bibinfo {author} {\bibfnamefont {C.~M.}\ \bibnamefont {Tonauer}}, \bibinfo {author} {\bibfnamefont {T.}~\bibnamefont {Loerting}}, \bibinfo {author} {\bibfnamefont {C.}~\bibnamefont {Gainaru}},\ and\ \bibinfo {author} {\bibfnamefont {R.}~\bibnamefont {Böhmer}},\ }\bibfield  {title} {\bibinfo {title} {Isotope effects on the dynamics of amorphous ices and aqueous phosphoric acid solutions},\ }\href {https://doi.org/10.1039/d2cp01455f} {\bibfield  {journal} {\bibinfo  {journal} {Physical Chemistry Chemical Physics}\ }\textbf {\bibinfo {volume} {24}},\ \bibinfo {pages} {14846} (\bibinfo {year} {2022})}\BibitemShut {NoStop}%
\bibitem [{\citenamefont {Hornig}\ \emph {et~al.}(1958)\citenamefont {Hornig}, \citenamefont {White},\ and\ \citenamefont {Reding}}]{Hor58}%
  \BibitemOpen
  \bibfield  {author} {\bibinfo {author} {\bibfnamefont {D.~F.}\ \bibnamefont {Hornig}}, \bibinfo {author} {\bibfnamefont {H.~F.}\ \bibnamefont {White}},\ and\ \bibinfo {author} {\bibfnamefont {F.~P.}\ \bibnamefont {Reding}},\ }\bibfield  {title} {\bibinfo {title} {The infrared spectrum of crystalline {H}$_2${O}, {D}$_2${O}, and {HDO}},\ }\href {https://doi.org/10.1016/0371-1951(58)80060-0} {\bibfield  {journal} {\bibinfo  {journal} {Spectrochimica Acta}\ }\textbf {\bibinfo {volume} {12}},\ \bibinfo {pages} {338} (\bibinfo {year} {1958})}\BibitemShut {NoStop}%
\bibitem [{\citenamefont {Petrenko}\ and\ \citenamefont {Whitworth}(1999)}]{Pet99}%
  \BibitemOpen
  \bibfield  {author} {\bibinfo {author} {\bibfnamefont {V.~F.}\ \bibnamefont {Petrenko}}\ and\ \bibinfo {author} {\bibfnamefont {R.~W.}\ \bibnamefont {Whitworth}},\ }\href@noop {} {\emph {\bibinfo {title} {Physics of Ice}}}\ (\bibinfo  {publisher} {Oxford University Press},\ \bibinfo {address} {Oxford},\ \bibinfo {year} {1999})\BibitemShut {NoStop}%
\bibitem [{\citenamefont {Artemov}(2019)}]{Art19}%
  \BibitemOpen
  \bibfield  {author} {\bibinfo {author} {\bibfnamefont {V.~G.}\ \bibnamefont {Artemov}},\ }\bibfield  {title} {\bibinfo {title} {A unified mechanism for ice and water electrical conductivity from direct current to terahertz},\ }\href {https://doi.org/https://doi.org/10.1039/C9CP00257J} {\bibfield  {journal} {\bibinfo  {journal} {Phys. Chem. Chem. Phys.}\ }\textbf {\bibinfo {volume} {21}},\ \bibinfo {pages} {8067} (\bibinfo {year} {2019})}\BibitemShut {NoStop}%
\bibitem [{\citenamefont {Hobbs}(1974)}]{Hob74}%
  \BibitemOpen
  \bibfield  {author} {\bibinfo {author} {\bibfnamefont {P.~V.}\ \bibnamefont {Hobbs}},\ }\href@noop {} {\emph {\bibinfo {title} {Ice Physics}}}\ (\bibinfo  {publisher} {Clarendon Press},\ \bibinfo {address} {Oxford},\ \bibinfo {year} {1974})\BibitemShut {NoStop}%
\bibitem [{\citenamefont {von Hippel}(1988)}]{Hip88}%
  \BibitemOpen
  \bibfield  {author} {\bibinfo {author} {\bibfnamefont {A.}~\bibnamefont {von Hippel}},\ }\bibfield  {title} {\bibinfo {title} {The dielectric relaxation spectra of water, ice, and aqueous solutions, and their interpretation: Critical survey of the status-quo for water},\ }\href@noop {} {\bibfield  {journal} {\bibinfo  {journal} {IEEE Transactions on Electrical Insulation}\ }\textbf {\bibinfo {volume} {23}},\ \bibinfo {pages} {801} (\bibinfo {year} {1988})}\BibitemShut {NoStop}%
\bibitem [{\citenamefont {Takei}(2007)}]{Tak07}%
  \BibitemOpen
  \bibfield  {author} {\bibinfo {author} {\bibfnamefont {I.}~\bibnamefont {Takei}},\ }\bibfield  {title} {\bibinfo {title} {Physics and chemistry of ice},\ }in\ \href@noop {} {\emph {\bibinfo {booktitle} {Proceedings of the 11th International Conference on the Physics and Chemistry of Ice}}},\ Vol.\ \bibinfo {volume} {430},\ \bibinfo {editor} {edited by\ \bibinfo {editor} {\bibfnamefont {W.~F.}\ \bibnamefont {Kuhs}}}\ (\bibinfo {year} {2007})\BibitemShut {NoStop}%
\bibitem [{\citenamefont {Warren}\ and\ \citenamefont {Brandt}(2008)}]{War08}%
  \BibitemOpen
  \bibfield  {author} {\bibinfo {author} {\bibfnamefont {S.~G.}\ \bibnamefont {Warren}}\ and\ \bibinfo {author} {\bibfnamefont {R.~E.}\ \bibnamefont {Brandt}},\ }\bibfield  {title} {\bibinfo {title} {Optical constants of ice from the ultraviolet to the microwave: a revised compilation},\ }\href@noop {} {\bibfield  {journal} {\bibinfo  {journal} {Journal of Geophysical Research: Atmospheres}\ }\textbf {\bibinfo {volume} {113}},\ \bibinfo {pages} {D14220} (\bibinfo {year} {2008})}\BibitemShut {NoStop}%
\bibitem [{\citenamefont {Matsuoka}\ \emph {et~al.}(1996)\citenamefont {Matsuoka}, \citenamefont {Fujita},\ and\ \citenamefont {Mae}}]{Mat96}%
  \BibitemOpen
  \bibfield  {author} {\bibinfo {author} {\bibfnamefont {T.}~\bibnamefont {Matsuoka}}, \bibinfo {author} {\bibfnamefont {S.}~\bibnamefont {Fujita}},\ and\ \bibinfo {author} {\bibfnamefont {S.}~\bibnamefont {Mae}},\ }\bibfield  {title} {\bibinfo {title} {Effect of temperature on dielectric properties of ice in the range 5–39 ghz},\ }\href@noop {} {\bibfield  {journal} {\bibinfo  {journal} {Journal of Applied Physics}\ }\textbf {\bibinfo {volume} {80}},\ \bibinfo {pages} {5884} (\bibinfo {year} {1996})}\BibitemShut {NoStop}%
\bibitem [{\citenamefont {Hobbs}\ \emph {et~al.}(1966)\citenamefont {Hobbs}, \citenamefont {Jhon},\ and\ \citenamefont {Eyring}}]{Hob66}%
  \BibitemOpen
  \bibfield  {author} {\bibinfo {author} {\bibfnamefont {M.~E.}\ \bibnamefont {Hobbs}}, \bibinfo {author} {\bibfnamefont {M.~S.}\ \bibnamefont {Jhon}},\ and\ \bibinfo {author} {\bibfnamefont {H.}~\bibnamefont {Eyring}},\ }\bibfield  {title} {\bibinfo {title} {The dielectric constant of liquid water and various forms of ice according to significant structure theory},\ }\href {https://doi.org/10.1073/pnas.56.1.31} {\bibfield  {journal} {\bibinfo  {journal} {Proceedings of the National Academy of Sciences}\ }\textbf {\bibinfo {volume} {56}},\ \bibinfo {pages} {31} (\bibinfo {year} {1966})}\BibitemShut {NoStop}%
\bibitem [{\citenamefont {Kornyshev}\ \emph {et~al.}(2003)\citenamefont {Kornyshev}, \citenamefont {Kuznetsov}, \citenamefont {Spohr},\ and\ \citenamefont {Ulstrup}}]{Kor03}%
  \BibitemOpen
  \bibfield  {author} {\bibinfo {author} {\bibfnamefont {A.~A.}\ \bibnamefont {Kornyshev}}, \bibinfo {author} {\bibfnamefont {A.~M.}\ \bibnamefont {Kuznetsov}}, \bibinfo {author} {\bibfnamefont {E.}~\bibnamefont {Spohr}},\ and\ \bibinfo {author} {\bibfnamefont {J.}~\bibnamefont {Ulstrup}},\ }\bibfield  {title} {\bibinfo {title} {Kinetics of proton transport in water},\ }\href {https://doi.org/10.1021/jp020857d} {\bibfield  {journal} {\bibinfo  {journal} {Journal of Physical Chemistry B}\ }\textbf {\bibinfo {volume} {107}},\ \bibinfo {pages} {3351} (\bibinfo {year} {2003})}\BibitemShut {NoStop}%
\bibitem [{\citenamefont {Artemov}(2021)}]{Art21}%
  \BibitemOpen
  \bibfield  {author} {\bibinfo {author} {\bibfnamefont {V.}~\bibnamefont {Artemov}},\ }\href@noop {} {\emph {\bibinfo {title} {The Electrodynamics of Water and Ice}}}\ (\bibinfo  {publisher} {Springer},\ \bibinfo {address} {Cham},\ \bibinfo {year} {2021})\BibitemShut {NoStop}%
\bibitem [{\citenamefont {Auty}\ and\ \citenamefont {Cole}(1952)}]{Aut52}%
  \BibitemOpen
  \bibfield  {author} {\bibinfo {author} {\bibfnamefont {R.~P.}\ \bibnamefont {Auty}}\ and\ \bibinfo {author} {\bibfnamefont {R.~H.}\ \bibnamefont {Cole}},\ }\bibfield  {title} {\bibinfo {title} {Dielectric properties of ice and solid {D}\textsubscript{2}{O}},\ }\href {https://doi.org/10.1063/1.1700726} {\bibfield  {journal} {\bibinfo  {journal} {J. Chem. Phys.}\ }\textbf {\bibinfo {volume} {20}},\ \bibinfo {pages} {1309} (\bibinfo {year} {1952})}\BibitemShut {NoStop}%
\bibitem [{\citenamefont {Johari}(1976)}]{Joh76}%
  \BibitemOpen
  \bibfield  {author} {\bibinfo {author} {\bibfnamefont {G.~P.}\ \bibnamefont {Johari}},\ }\bibfield  {title} {\bibinfo {title} {The dielectric properties of {H}\textsubscript{2}{O} and {D}\textsubscript{2}{O} ice {I}h at {M}{H}z frequencies},\ }\href {https://doi.org/10.1063/1.432033} {\bibfield  {journal} {\bibinfo  {journal} {The Journal of Chemical Physics}\ }\textbf {\bibinfo {volume} {64}},\ \bibinfo {pages} {3998} (\bibinfo {year} {1976})}\BibitemShut {NoStop}%
\bibitem [{\citenamefont {Kawada}(1979)}]{Kaw79}%
  \BibitemOpen
  \bibfield  {author} {\bibinfo {author} {\bibfnamefont {S.}~\bibnamefont {Kawada}},\ }\bibfield  {title} {\bibinfo {title} {Dielectric properties of heavy ice {I}h ({D}\textsubscript{2}{O} ice)},\ }\href {https://doi.org/10.1143/JPSJ.47.1850} {\bibfield  {journal} {\bibinfo  {journal} {Journal of the Physical Society of Japan}\ }\textbf {\bibinfo {volume} {47}},\ \bibinfo {pages} {1850} (\bibinfo {year} {1979})}\BibitemShut {NoStop}%
\bibitem [{\citenamefont {Okada}\ \emph {et~al.}(1999)\citenamefont {Okada}, \citenamefont {Yao}, \citenamefont {Hiejima}, \citenamefont {Kohno},\ and\ \citenamefont {Kajihara}}]{Oka99}%
  \BibitemOpen
  \bibfield  {author} {\bibinfo {author} {\bibfnamefont {K.}~\bibnamefont {Okada}}, \bibinfo {author} {\bibfnamefont {M.}~\bibnamefont {Yao}}, \bibinfo {author} {\bibfnamefont {Y.}~\bibnamefont {Hiejima}}, \bibinfo {author} {\bibfnamefont {H.}~\bibnamefont {Kohno}},\ and\ \bibinfo {author} {\bibfnamefont {Y.}~\bibnamefont {Kajihara}},\ }\bibfield  {title} {\bibinfo {title} {Dielectric relaxation of water and heavy water in the whole fluid phase},\ }\href {https://doi.org/10.1063/1.477897} {\bibfield  {journal} {\bibinfo  {journal} {Journal of Chemical Physics}\ }\textbf {\bibinfo {volume} {110}},\ \bibinfo {pages} {3026} (\bibinfo {year} {1999})}\BibitemShut {NoStop}%
\bibitem [{\citenamefont {Johari}\ and\ \citenamefont {Whalley}(1981)}]{Joh81}%
  \BibitemOpen
  \bibfield  {author} {\bibinfo {author} {\bibfnamefont {G.~P.}\ \bibnamefont {Johari}}\ and\ \bibinfo {author} {\bibfnamefont {E.}~\bibnamefont {Whalley}},\ }\bibfield  {title} {\bibinfo {title} {The dielectric properties of ice {I}h in the range 272–133 {K}},\ }\href {https://doi.org/10.1063/1.442139} {\bibfield  {journal} {\bibinfo  {journal} {J. Chem. Phys.}\ }\textbf {\bibinfo {volume} {75}},\ \bibinfo {pages} {1333} (\bibinfo {year} {1981})}\BibitemShut {NoStop}%
\bibitem [{\citenamefont {Johari}\ and\ \citenamefont {Jones}(1978)}]{Joh78}%
  \BibitemOpen
  \bibfield  {author} {\bibinfo {author} {\bibfnamefont {G.~P.}\ \bibnamefont {Johari}}\ and\ \bibinfo {author} {\bibfnamefont {S.~J.}\ \bibnamefont {Jones}},\ }\bibfield  {title} {\bibinfo {title} {The orientation polarization in hexagonal ice parallel and perpendicular to the c-axis},\ }\href {https://doi.org/10.1017/S0022143000033463} {\bibfield  {journal} {\bibinfo  {journal} {Journal of Glaciology}\ }\textbf {\bibinfo {volume} {21}},\ \bibinfo {pages} {259} (\bibinfo {year} {1978})}\BibitemShut {NoStop}%
\bibitem [{\citenamefont {Bruni}\ \emph {et~al.}(1993)\citenamefont {Bruni}, \citenamefont {Consolini},\ and\ \citenamefont {Careri}}]{Bru93}%
  \BibitemOpen
  \bibfield  {author} {\bibinfo {author} {\bibfnamefont {F.}~\bibnamefont {Bruni}}, \bibinfo {author} {\bibfnamefont {G.}~\bibnamefont {Consolini}},\ and\ \bibinfo {author} {\bibfnamefont {G.}~\bibnamefont {Careri}},\ }\bibfield  {title} {\bibinfo {title} {Temperature dependence of dielectric relaxation in {H}\textsubscript{2}{O} and {D}\textsubscript{2}{O} ice. a dissipative quantum tunneling approach},\ }\href {https://doi.org/10.1063/1.465778} {\bibfield  {journal} {\bibinfo  {journal} {J. Chem. Phys.}\ }\textbf {\bibinfo {volume} {99}},\ \bibinfo {pages} {538} (\bibinfo {year} {1993})}\BibitemShut {NoStop}%
\bibitem [{\citenamefont {Kremer}\ and\ \citenamefont {Schönhals}(2003)}]{Kre03}%
  \BibitemOpen
  \bibinfo {editor} {\bibfnamefont {F.}~\bibnamefont {Kremer}}\ and\ \bibinfo {editor} {\bibfnamefont {A.}~\bibnamefont {Schönhals}},\ eds.,\ \href {https://doi.org/10.1007/978-3-642-56120-7} {\emph {\bibinfo {title} {Broadband Dielectric Spectroscopy}}},\ \bibinfo {edition} {1st}\ ed.\ (\bibinfo  {publisher} {Springer},\ \bibinfo {address} {Berlin},\ \bibinfo {year} {2003})\ pp.\ \bibinfo {pages} {XXI, 729}\BibitemShut {NoStop}%
\bibitem [{\citenamefont {Artemov}\ and\ \citenamefont {Volkov}(2014)}]{Art14}%
  \BibitemOpen
  \bibfield  {author} {\bibinfo {author} {\bibfnamefont {V.~G.}\ \bibnamefont {Artemov}}\ and\ \bibinfo {author} {\bibfnamefont {A.~A.}\ \bibnamefont {Volkov}},\ }\bibfield  {title} {\bibinfo {title} {Water and ice dielectric spectra scaling at 0\textdegree{}{C}},\ }\href {https://doi.org/10.1080/00150193.2014.895216} {\bibfield  {journal} {\bibinfo  {journal} {Ferroelectrics}\ }\textbf {\bibinfo {volume} {466}},\ \bibinfo {pages} {158} (\bibinfo {year} {2014})}\BibitemShut {NoStop}%
\bibitem [{\citenamefont {Kramers}(1940)}]{Kra40}%
  \BibitemOpen
  \bibfield  {author} {\bibinfo {author} {\bibfnamefont {H.~A.}\ \bibnamefont {Kramers}},\ }\bibfield  {title} {\bibinfo {title} {Brownian motion in a field of force and the diffusion model of chemical reactions},\ }\href {https://doi.org/10.1016/S0031-8914(40)90098-2} {\bibfield  {journal} {\bibinfo  {journal} {Physica}\ }\textbf {\bibinfo {volume} {7}},\ \bibinfo {pages} {284} (\bibinfo {year} {1940})}\BibitemShut {NoStop}%
\bibitem [{\citenamefont {H{\"a}nggi}\ \emph {et~al.}(1990)\citenamefont {H{\"a}nggi}, \citenamefont {Talkner},\ and\ \citenamefont {Borkovec}}]{Han90}%
  \BibitemOpen
  \bibfield  {author} {\bibinfo {author} {\bibfnamefont {P.}~\bibnamefont {H{\"a}nggi}}, \bibinfo {author} {\bibfnamefont {P.}~\bibnamefont {Talkner}},\ and\ \bibinfo {author} {\bibfnamefont {M.}~\bibnamefont {Borkovec}},\ }\bibfield  {title} {\bibinfo {title} {Reaction-rate theory: fifty years after kramers},\ }\href {https://doi.org/10.1103/RevModPhys.62.251} {\bibfield  {journal} {\bibinfo  {journal} {Reviews of Modern Physics}\ }\textbf {\bibinfo {volume} {62}},\ \bibinfo {pages} {251} (\bibinfo {year} {1990})}\BibitemShut {NoStop}%
\bibitem [{Note1()}]{Note1}%
  \BibitemOpen
  \bibinfo {note} {Each O–O bond hosts exactly one proton, displaced from the midpoint so that each oxygen atom is coordinated by two near and two distant protons, giving rise to residual entropy.}\BibitemShut {Stop}%
\bibitem [{Note2()}]{Note2}%
  \BibitemOpen
  \bibinfo {note} {Bjerrum defects are a bond without a proton (L-defect) or with two protons (D-defect).}\BibitemShut {Stop}%
\bibitem [{Note3()}]{Note3}%
  \BibitemOpen
  \bibinfo {note} {Bjerrum argued [Bjerrum, N. Untersuchungen über Ionenassoziation I. K. Dan. Vidensk. Selsk. 1926, 7, 1–48] that under normal conditions thermal energy is insufficient to overcome Coulomb attraction, so ions should form pairs rather than remain free.}\BibitemShut {Stop}%
\bibitem [{\citenamefont {Geissler}\ \emph {et~al.}(2001)\citenamefont {Geissler}, \citenamefont {Dellago}, \citenamefont {Chandler}, \citenamefont {Hutter},\ and\ \citenamefont {Parrinello}}]{Gei01}%
  \BibitemOpen
  \bibfield  {author} {\bibinfo {author} {\bibfnamefont {P.~L.}\ \bibnamefont {Geissler}}, \bibinfo {author} {\bibfnamefont {C.}~\bibnamefont {Dellago}}, \bibinfo {author} {\bibfnamefont {D.}~\bibnamefont {Chandler}}, \bibinfo {author} {\bibfnamefont {J.}~\bibnamefont {Hutter}},\ and\ \bibinfo {author} {\bibfnamefont {M.}~\bibnamefont {Parrinello}},\ }\bibfield  {title} {\bibinfo {title} {Autoionization in liquid water},\ }\href {https://doi.org/10.1126/science.1056991} {\bibfield  {journal} {\bibinfo  {journal} {Science}\ }\textbf {\bibinfo {volume} {291}},\ \bibinfo {pages} {2121} (\bibinfo {year} {2001})}\BibitemShut {NoStop}%
\bibitem [{\citenamefont {Ceriotti}\ \emph {et~al.}(2013)\citenamefont {Ceriotti}, \citenamefont {Cuny}, \citenamefont {Parrinello},\ and\ \citenamefont {Manolopoulos}}]{Cer13}%
  \BibitemOpen
  \bibfield  {author} {\bibinfo {author} {\bibfnamefont {M.}~\bibnamefont {Ceriotti}}, \bibinfo {author} {\bibfnamefont {J.}~\bibnamefont {Cuny}}, \bibinfo {author} {\bibfnamefont {M.}~\bibnamefont {Parrinello}},\ and\ \bibinfo {author} {\bibfnamefont {D.~E.}\ \bibnamefont {Manolopoulos}},\ }\bibfield  {title} {\bibinfo {title} {Nuclear quantum effects and hydrogen bond fluctuations in water},\ }\href {https://doi.org/10.1073/pnas.1308560110} {\bibfield  {journal} {\bibinfo  {journal} {Proceedings of the National Academy of Sciences}\ }\textbf {\bibinfo {volume} {110}},\ \bibinfo {pages} {15591} (\bibinfo {year} {2013})}\BibitemShut {NoStop}%
\bibitem [{\citenamefont {Onsager}(1936)}]{Ons36}%
  \BibitemOpen
  \bibfield  {author} {\bibinfo {author} {\bibfnamefont {L.}~\bibnamefont {Onsager}},\ }\bibfield  {title} {\bibinfo {title} {Electric moments of molecules in liquids},\ }\href {https://doi.org/10.1021/ja01299a050} {\bibfield  {journal} {\bibinfo  {journal} {Journal of the American Chemical Society}\ }\textbf {\bibinfo {volume} {58}},\ \bibinfo {pages} {1486} (\bibinfo {year} {1936})}\BibitemShut {NoStop}%
\bibitem [{\citenamefont {Fletcher}(1987)}]{Fle87}%
  \BibitemOpen
  \bibfield  {author} {\bibinfo {author} {\bibfnamefont {R.}~\bibnamefont {Fletcher}},\ }\href@noop {} {\emph {\bibinfo {title} {Practical Methods of Optimization}}},\ \bibinfo {edition} {2nd}\ ed.\ (\bibinfo  {publisher} {John Wiley \& Sons},\ \bibinfo {address} {New York},\ \bibinfo {year} {1987})\BibitemShut {NoStop}%
\bibitem [{\citenamefont {Landau}\ and\ \citenamefont {Lifshitz}(1980)}]{Lan80}%
  \BibitemOpen
  \bibfield  {author} {\bibinfo {author} {\bibfnamefont {L.~D.}\ \bibnamefont {Landau}}\ and\ \bibinfo {author} {\bibfnamefont {E.~M.}\ \bibnamefont {Lifshitz}},\ }\href@noop {} {\emph {\bibinfo {title} {Statistical Physics, Part 1}}},\ \bibinfo {edition} {3rd}\ ed.,\ \bibinfo {series} {Course of Theoretical Physics}, Vol.~\bibinfo {volume} {5}\ (\bibinfo  {publisher} {Butterworth-Heinemann},\ \bibinfo {address} {Oxford, England},\ \bibinfo {year} {1980})\BibitemShut {NoStop}%
\end{thebibliography}%

\end{document}